\documentclass{article}

\usepackage{amssymb,amsfonts,amsmath,stmaryrd}
\usepackage{cite,enumerate,float,indentfirst}
\usepackage{color}
\usepackage{empheq}
\usepackage{url}
\usepackage{hyperref}
\usepackage{textcomp}

\def\be{\begin{eqnarray}}
\def\ee{\end{eqnarray}}

\def\Tr{{\rm Tr}\,}

\newcommand\Rbracket[1]{\left(#1\right)}
\newcommand\Sbracket[1]{\left[#1\right]}

\definecolor{red}{rgb}{1,0,0}
\definecolor{orange}{rgb}{1,0.5,0}
\definecolor{violet}{rgb}{0.7,0,1}

\hoffset= - 1.0in         

\def\boxSize{20}
\def\boxPic{
  \put(0,0){
    \put(0,0){\line(0,-1){\boxSize}}
    \put(0,0){\line(1,0){\boxSize}}
    \put(\boxSize,0){\line(0,-1){\boxSize}}
    \put(0,-\boxSize){\line(1,0){\boxSize}}
  }
}

\begin{document}

\title{\vspace{-1cm}{\Large {\bf
      Miwa deformation of WLZZ matrix models
    }}}
\date{}
\author{T.Diudin$^{a}$, A.Popolitov$^{a,b,c}$}

\maketitle
\vspace{-4.2cm}

\begin{center}
	\hfill ITEP/TH-35/26 \\
	\hfill IITP/TH-29/26 \\
	\hfill MIPT/TH-27/26
\end{center}

\vspace{1.7cm}

\begin{center}
$^a$ {\small {\it MIPT, Dolgoprudny, 141701, Russia}}\\
$^b$ {\small {\it NRC ``Kurchatov Institute", 123182, Moscow, Russia}}\\
$^c$ {\small {\it Institute for Information Transmission Problems, Moscow 127994, Russia}}
\end{center}

\vspace{0.5cm}

\begin{abstract}
  In this paper we propose a universal way to define multivariate orthogonal polynomials
  for the positive branch of WLZZ matrix models.
  This allows us to generalize from the Gaussian case, previously
  considered in the literature, the strict superintegrability property and
  the Miwa deformation formulas.
\end{abstract}

\bigskip

\section{Introduction}\label{sec:introduction}

Matrix models \cite{paper:Mehta} are ubiquitous in modern mathematical physics, with connections to broadest range of
topics: from random Markov processes \cite{paper:Dyson}
, to topological 2D quantum gravity \cite{paper:Witten-2d-top}, to
integrability \cite{paper:M-mm-as-integrable-systems}, enumerative geometry
\cite{paper:Alexandrov-mmrp}
and supersymmetric Yang-Mills-Chern-Simons theories\cite{paper:CZ-ymcs}. Understood originally as \textit{literal}
Riemann integrals over certain space of matrices, matrix models were quickly understood to
possess plethora of \textit{emergent properties} (hidden symmetries) -- and research focus
shifted accordingly.

One of such approaches, which recently gained much research attention, links certain matrix model
partition functions to representation theory of $W_{1+\infty}$ algebra and in more complicated
cases the affine Yangian $Y(gl_1)$ and DIM algebra. The relevant class of matrix models are the
WLZZ matrix models \cite{paper:WLZZ-original,paper:MM-wlzz-int-sys}.

In particular, this link to algebra allows to conveniently prove another mysterious emergent
property which some matrix models possess -- the superintegrability
\cite{paper:MM-super-summ}. Superintegrability of a matrix model implies existence of a superintegrable basis of polynomials. Averages of these polynomials are equal to said polynomials evaluated at certain, model dependent, special point

\begin{align}
    \forall \Delta: \  <P_{\Delta}> \ = \ P_{\Delta}(*)
\end{align}

For more common and more simple models their suitable systems of polynomials $P_{\Delta}$ are known (for instance, in Hermitian Gaussian matrix model or HGMM which we recall in
Section \ref{sec:gaussian-case} the basis is Schur polynomials $S_{\Delta}$ indexed by partitions $\Delta$).

\bigskip

Furthermore, for some models there exists an even stronger property called 
\textit{strict superintegrability}\cite{paper:MM-strict-superint}. It amounts to existence of a new system of polynomials $K_\Delta$ so that certain {\bf double} averages are equally apparent:
\begin{align} \label{eq:definition-of-K}
    <P_{\Delta}K_{\Delta'}> \ \sim \ P_{\Delta / \Delta'}(*)
\end{align}
where new polynomials are orthogonalizing the native model average:
\begin{align}
    <K_{\Delta}K_{\Delta'}> \ \sim \ \delta_{\Delta, \Delta'}
\end{align}
making $K_{\Delta}$ multivariate orthogonal polynomials for this model.

\bigskip

It is quite tedious to calculate $K_{\Delta}$ directly from their definition \eqref{eq:definition-of-K} but in \cite{paper:MMM-single-equation} they were shown to be solution to {\bf dual Single equation}. The problem, however, is that $K_{\Delta}$ is not unique solution of dual Single equation and therefore a question of postulating the "correct" distinguished solution remains open (see Section \ref{sec:gaussian-case} for details). In known examples the right solution is singled out by postulating
that highest degree of $K_\Delta$ is Schur polynomial $S_\Delta$, however, this requirement is
\textit{ad hoc} and it is not clear what should replace it in general. It is therefore natural to search for additional constricting equations for $K_{\Delta}$ which together with the dual Single equation would make the distinguished solution unique.

\bigskip

In this paper we argue that the source for the additional equations may be provided by
abovementioned link to $W_{1+\infty}$ algebra. Namely, for a given WLZZ matrix model other
equations are given by action of other Hamiltonians on the same commuting ray. These extra equations have kind of {\bf ladder} shape (see Section \ref{sec:tow-inv-kdelta}), while the original dual Single equation is an eigenequation.

\bigskip

Interestingly, both the ladder equations and the eigenequation are necessary to uniquely define $K_{\Delta}$ as we illustrate in Section \ref{sec:study-of-freedom} with Gaussian model example. If one considers only the ladder equations, discarding the dual Single equation, there is in fact quite large freedom in solution space.

\bigskip

In Section \ref{sec:wlzz} our approach allows us to obtain $K_\Delta$ polynomials for any
WLZZ model on the positive branch\footnote{Treatment of the negative branch is analogous but
requires more care due to extra exponential term} and hence obtain corresponding Miwa-deformations
for these models

To summarize, in case of (positive branch of) WLZZ models our prescription works in the most
straightforward way, and we conclude (Section~\ref{sec:conclusion}) with a list of more
complicated setups where it can be tested further and hopefully lead to new insights.


\section{The Gaussian case} \label{sec:gaussian-case}
In this section we review superintegrability, strict superintegrability and Miwa-deformation of HGMM following \cite{paper:MMPSh-miwa}.

\bigskip

The partition function of Hermitian Gaussian matrix model with times $p_k$ can be defined as
\begin{align} \label{eq:partition-function-in-HGMM}
Z_2(p) \ = & \ \int DX \exp{\Rbracket{-\frac{1}{2} \Tr X^2 }} \cdot \exp{\Rbracket{\overset{\infty}{\underset{k=1}{\sum}} \frac{p_k}{k} \Tr X^k}}
\end{align}
From the point of view of WLZZ two-dimensional family of models, this partition function is $Z^{(1)}_2(p)$, $Z^{(n)}_m$ being the general function. The integral in \eqref{eq:partition-function-in-HGMM} is understood as a formal power series in times $p_k$. By evaluating derivatives at origin one may calculate any multi-trace correlator:
\begin{align} \label{eq:averages-in-HGMM}
<\Tr X^{k_1} \ldots \Tr X^{k_m}>_2 & \ = \ 
\left(\frac{k_1 \partial}{\partial p_{k_1}} \ldots \frac{k_n \partial}{\partial p_{k_n}} Z_2
\right)\Bigg{|}_{p=0}
\end{align}
\noindent Via insertion of suitably chosen full derivative operators under the integral, one may arrive at Virasoro constraints for this model \cite{paper:M-virasoro}:
\begin{align}
\Sbracket{\widehat{\mathcal{L}}^{vac}_n - (n+2)\frac{\partial}{\partial p_{n+2}}} Z_2 \ =  \ 0
\end{align}
\begin{align}
  \widehat{\mathcal{L}}^{vac}_n \ = \ \underset{l}{\sum} \Rbracket{(l+n) p_l\frac{\partial}{\partial p_{l+n}}} + \overset{n-1}{\underset{s=1}{\sum}} \Rbracket{s(n-s) \Sbracket{\frac{\partial}{\partial p_s} \frac{\partial}{\partial p_{n-s}}}} + 2Nn \frac{\partial}{\partial p_{n}} + N^2 \delta_{n, 0} + Np_1 \delta_{n+1, 0}
\end{align}
which can be summed up to the so-called Single equation \cite{paper:MMM-single-equation}:
\begin{align} \label{eq:gauss-single-equation}
  (D-W_{-2}) \cdot Z_2 = 0
\end{align}
\begin{align}
D \ = & \ \overset{\infty}{\underset{k=1}{\sum}} kp_k \frac{\partial}{\partial p_k}
\end{align}

Here $W_{-2}$ is the celebrated "W-" (or cut-and-join) operator \cite{paper:MSh-w-ops}:
\begin{align}
  W_{-2} \ = & \ \overset{\infty}{\underset{b=0}{\sum}}  \overset{\infty}{\underset{a=0}{\sum}} \Sbracket{ (a+b-2) p_a p_b \frac{\partial}{\partial p_{a+b-2}} + a b  p_{a+b+2} \frac{\partial}{\partial p_{a}} \frac{\partial}{\partial p_{b}}} + (N p_1^2 + N^2 p_2)
\end{align}

Thanks to BCH formula, Single equation \eqref{eq:gauss-single-equation} has unique solution which is a formal power series in times $p_k$, up to normalization
\begin{align}
    Z_2 \ = & \ \Rbracket{\exp{\frac{W_{-2}}{2}}} \cdot 1
\end{align}

Following \cite{paper:CM-quant-rec}, to construct multipvariate orthogonal polynomials one is to consider eigenequation dual \footnote{Duality here is understood w.r.t Macdonald scalar product
$<p_k|p_l> = k \delta_{k,l}$, i.e.
$p_k \leftrightarrow \frac{\partial}{\partial p_k}$}
to the Single equation \eqref{eq:gauss-single-equation} -- the \textbf{dual Single equation}:
\begin{align} \label{eq:gauss-dual-single-equation}
  (D + W_2) \cdot Q_\Delta = & \ |\Delta| Q_\Delta
\end{align}
\begin{align}
  W_2 \ = & \ \overset{\infty}{\underset{a,b=0}{\sum}} \Rbracket{a b p_{a+b-2} \frac{\partial}{\partial p_a} \frac{\partial}{\partial p_b} + (a+b+2) p_a p_b \frac{\partial}{\partial p_{a+b+2}}}
\end{align}
where the right hand side is generalized to include non-trivial
eigenvalue $\lambda = |\Delta|$. Since the dual cut-and-join operator $W_2$ lowers the grading \footnote{We recall that $p_k$ has grading k and product $\underset{i}{\prod} p_i$ has grading $\underset{i}{\sum} i$} by 2, it seems natural to search for {\bf polynomial} solutions to \eqref{eq:gauss-dual-single-equation}. Even in the class of polynomials the solution is not unique, but there is a distinguished solution $Q_{\Delta} = K_{\Delta}$:
\begin{align} \label{eq:hermite-poly-w-gauss}
  K_\Delta \ = & \ \Rbracket{\exp{\frac{W_2}{2}}} \cdot S_\Delta
\end{align}
Its highest degree (w.r.t grading operator $D$) is Schur polynomial $S_{\Delta}$. A somewhat reasonable justification for this requirement is that Schur polynomials form superintegrable basis for HGMM. This can be used to easily calculate averages \footnote{Here $\eta_{\Delta}(N) = \eta(\Delta) = \frac{S_{\Delta}(N)}{S_{\Delta}(\delta_{k, 1})} = \underset{(i, j) \in \Delta}{\prod} (N+j-i)$} :
\begin{align} \label{eq:gauss-superint}
 <S_\Delta (p_k)>_2 \ = & \ S_\Delta(\delta_{k, 2}) \eta_{\Delta}(N);
 \ \ \eta_{\Delta}(N) = \prod_{(i,j)\in\Delta}\left(N - i + j\right)
\end{align}

This particular justification, however, cannot work in situations where one does not know the superintegrable basis. In fact, one of the motivations for the present project is to obtain such unambiguous definition of $K_{\Delta}$ from the model that it in turn allows to find the superintegrable basis as the highest degree of the $K_{\Delta}$ polynomials.

\bigskip

In this Gaussian case the distinguished solution \eqref{eq:hermite-poly-w-gauss} is, in fact, nothing but the multivariate Hermite polynomial, which has a number of equivalent definitions:
\begin{align}
K_{\Delta}\left\{p_k=\Tr X^k\right\} = \exp{\Rbracket{\frac{1}{2} \Tr X^2}} S_{\Delta} \Rbracket{p_k = Tr \Rbracket{\frac{\partial}{\partial X}}^k} \exp{\Rbracket{-\frac{1}{2} \Tr X^2}}
\end{align}
\begin{align}\label{eq:eq-of-K-gauss}
K_{\Delta}(p_k) = \underset{\Delta' \subseteq \Delta}{\sum} S_{\Delta'} (p_k) S_{\Delta / \Delta'}(\delta_{k,2}) \frac{\eta_{\Delta}(N)}{\eta_{\Delta'}(N)}
\end{align}
and its importance to the Hermitian Gaussian ensemble is that, first of all,
double average of these polynomials is simple and diagonal
\begin{align}\label{eq:diagonality-gauss}
<K_{\Delta} \cdot K _{\Delta'}>_2 \ = & \ S_{\Delta}(N) \delta_{\Delta, \Delta'}
\end{align}
which is to be expected from multivariate orthogonal polynomials \cite{paper:CM-quant-rec}.

\bigskip

Far less automatic and straightforward  is another double average, one that does hold in the case of HGMM, and is known in few other cases, but not in general -- the strict superintegrability property \cite{paper:MM-strict-superint}
\begin{align}\label{eq:strict-superint-gauss}
<S_{\Delta} \cdot K_{\Delta'}>_2 \ = & \ S_{\Delta / \Delta'}(\delta_{k, 2}) \eta_{\Delta}(N)
\end{align}

Importance of this strict superintegrability property in relation to Miwa deformation for
the model \cite{paper:MMPSh-miwa} is that given the Miwa-deformed partition function
\begin{align} \label{eq:partition-func-in-Miwa-deformed-HGMM}
Z_{2, \text{Miwa}}^{(1)} = & \ \underset{\Delta}{\sum} S_{\Delta}(p_k + \pi_k) \cdot S_{\Delta} (\delta_{k, 2}) \cdot \eta_\Delta(N) = \underset{\Delta, \Delta'}{\sum} S_{\Delta'}(p_k) \cdot S_{\Delta / \Delta'} (\pi_k) \cdot S_{\Delta} (\delta_{k, 2}) \cdot \eta_\Delta(N)
\\ \notag
=& \Bigg{|}_{\pi_k=\sum_{i=1}^M z_i^k} \ \int DX \exp{\Rbracket{-\frac{1}{2} \Tr X^2 }}
\cdot \frac{\exp{\Rbracket{\overset{\infty}{\underset{k=1}{\sum}} \frac{p_k}{k} \Tr X^k}}}{\prod_{i=1}^M \text{det}\left(1 - z_i X\right)}
\end{align}
the $K_\Delta$ averages are concisely and manifestly expressed
\begin{align} \label{eq:averages-in-Miwa-deformed-HGMM}
<K(\Delta')>_{\{s\}} = & \ \underset{\Delta}{\sum} S_{\Delta}(z_a) <S_{\Delta} \cdot K_{\Delta'}> = \underset{\Delta}{\sum} S_{\Delta}(z_a) S_{\Delta / \Delta'} (\delta_{k,2}) \cdot \eta_\Delta(N)
\\ \notag
= & \ \mathcal{B}_N \Sbracket{\underset{\Delta}{\sum} S_{\Delta}(z_a) \cdot S_{\Delta / \Delta'}(\delta_{k,2}) } = \mathcal{B}_N \Sbracket{S_{\Delta'}\{\underline{z}\}
e^{\frac{1}{2}\sum_a z_a^2}}
\end{align}
through multivariate analog of the Borel transform $\mathcal{B}_N$.

\bigskip

In following sections we show that these remarkable properties, both strict superintegrability and manifestation of Miwa-deformed averages generalize to WLZZ-models. First we need a suitable general definition for multivariate orthogonal polynomials $K_{\Delta}$.

\section{(Towards) invariant definition of $K_\Delta$ polynomials}
\label{sec:tow-inv-kdelta}

The choice of a particular solution \eqref{eq:hermite-poly-w-gauss} from many possible
solutions of \eqref{eq:gauss-dual-single-equation} may seem contrived, even though
we know a posteriori that it has very strong property \eqref{eq:strict-superint-gauss}
that is, in turn, linked to superintegrability of Miwa deformed HGMM
\cite{paper:MMPSh-miwa}.
In this section we try to make a choice \eqref{eq:hermite-poly-w-gauss} more natural.

\bigskip

Namely, the idea is to augment the dual Single equation by a consistent set of
differential equations in such a way that the distinguished solution \eqref{eq:hermite-poly-w-gauss}
is \textit{the only} solution of this system corresponding to our eigenequations.
A natural source of extra equations for such a system is provided by higher Hamiltonians
living on the same ray (in Yangian ($W_{1+\infty}$) algebra) as operator $W_2$ used in dual Single equation \cite{paper:MMM-comm-fams}.

\bigskip

In Schur basis, these higher Hamiltonians act as follows
\begin{align}
H_k(S_\Delta)=\underset{\Delta' = \Delta \backslash hook(k)}{\sum}  (-1)^{||hook(k)||} \cdot \frac{\eta_\Delta (N)}{\eta_{\Delta'} (N)} \cdot S_{\Delta'}.
\end{align}
where summation goes over all partitions one can acquire by removing \textit{k-hooks} from $\Delta$. We define  \textit{k-hook} (also known as k-border strip) as a continuous line  of $k$ "boxes" where every two consequtive boxes have common side. The the resulting set
of boxes $\Delta'$ should also be valid Young diagram, see Eqn.~\eqref{fig:k-hook}
for examples. The norm of a k-hook is the number of horizontal lines it spans,
for instance, correct hooks on Eqn.~\eqref{fig:k-hook} have norms equal 2 and 3, respectively.

  \begin{align} \label{fig:k-hook}
    \begin{picture}(300,130)(0,-130)
      \thicklines
      \def\boxSize{10}
\put(0,0){
    \put(0,0){
        \put(0,0){\boxPic}
        \put(10,0){\boxPic}
        \put(20,0){\boxPic}
        \put(30,0){\boxPic}
        \put(40,0){\boxPic}
        \put(0,-10){\boxPic}
        \put(10,-10){\boxPic}
        \put(20,-10){\boxPic}
        \put(0,-20){\boxPic}
        \put(10,-20){\boxPic}
        \put(25,-5){\color{blue}\line(1,0){20}}
        \put(25,-5){\color{blue}\line(0,-1){10}}
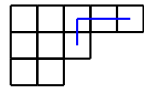
        \put(0,-50){$\substack{\text{Correct 4-hook} \\ \text{$||$hook$||$=2}}$}
    }
}
\put(150,0){
    \put(0,0){
        \put(0,0){\boxPic}
        \put(10,0){\boxPic}
        \put(20,0){\boxPic}
        \put(30,0){\boxPic}
        \put(40,0){\boxPic}
        \put(0,-10){\boxPic}
        \put(10,-10){\boxPic}
        \put(20,-10){\boxPic}
        \put(0,-20){\boxPic}
        \put(10,-20){\boxPic}
        \put(25,-5){\color{blue}\line(1,0){20}}
        \put(25,-5){\color{blue}\line(0,-1){10}}
        \put(15,-15){\color{blue}\line(1,0){10}}
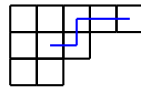
        \put(0,-50){$\substack{\text{Incorrect 5-hook} \\ \text{result not a partition}}$}
    }
}
\put(300,0){
    \put(0,0){
        \put(0,0){\boxPic}
        \put(10,0){\boxPic}
        \put(20,0){\boxPic}
        \put(30,0){\boxPic}
        \put(40,0){\boxPic}
        \put(0,-10){\boxPic}
        \put(10,-10){\boxPic}
        \put(20,-10){\boxPic}
        \put(0,-20){\boxPic}
        \put(10,-20){\boxPic}
        \put(25,-5){\color{blue}\line(1,0){20}}
        \put(25,-5){\color{blue}\line(0,-1){10}}
        \put(15,-15){\color{blue}\line(1,0){10}}
        \put(15,-15){\color{blue}\line(0,-1){10}}
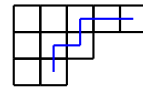
        \put(0,-50){$\substack{\text{Correct 6-hook} \\ \text{$||$hook$||$=3}}$}
    }
}
    \end{picture}
  \end{align}

It is straightforward to observe that the distinguished solution \eqref{eq:eq-of-K-gauss} satisfies similar constraints:
\begin{align} \label{eq:single-system-gauss}
H_k(K_\Delta)=\underset{\Delta' = \Delta - (k \ hook)}{\sum}  (-1)^{||k \ hook||} \cdot \frac{\eta_\Delta (N)}{\eta_{\Delta'} (N)} \cdot K_{\Delta'}
\end{align}

\bigskip

In \textbf{Appendix B} we directly verify that the solution to \eqref{eq:single-system-gauss} with additional constraint \eqref{eq:gauss-dual-single-equation} is, indeed, unique and given by \eqref{eq:hermite-poly-w-gauss}. Hence \eqref{eq:gauss-dual-single-equation} + \eqref{eq:single-system-gauss} may be used to invariantly define polynomials $K_\Delta$.

\bigskip

A reverse kind of question now naturally arises. Suppose we keep the "ladder" Hamiltonian equations \eqref{eq:single-system-gauss}, but drop the dual Single equation \eqref{eq:gauss-dual-single-equation}. Clearly then solution will no longer be unique (at the very least both ordinary Schur polynomials $S_{\Delta}$ and multivariate Hermite polynomials $K_{\Delta}$ \eqref{eq:eq-of-K-gauss} satisfy the "ladder" system). What will be the general form of its solution? This is what we analyze in the next section.

\section{Freedom in additional equations} \label{sec:study-of-freedom}
Now that we know that both $S_\Delta$ and $K_\Delta$ are solutions for \eqref{eq:single-system-gauss}, it may be reasonable to ask how the general solution of this system looks. We gradually construct polynomials $P_\Delta$, going layer by layer in $|\Delta|$. Note that since $H_k$ lowers the grading by $k$, on level $|\Delta|$ it is sufficient to consider action of first $|\Delta|$ Hamiltonians -- higher equations vanish.

\bigskip

First of all, without loss of generality we normalize the only \textbf{level 0} polynomial to 1:
\[
P_{\varnothing}=1
\]

On \textbf{layer 1} there is one polynomial subject to one equation:
\[
P_{[1]} = B_{[1]} p_1 + A_{[1]}, \ \Rightarrow \ B_{[1]} N = P_{\varnothing}\cdot N, \ \Rightarrow B_{[1]} = 1
\]
Therefore we get $P_{[1]} = S_{[1]} + A_{[1]}$: there is one free parameter $A_{[1]}$.
If we then also impose \eqref{eq:gauss-dual-single-equation} for $\Delta = [1]$ we get $A_{[1]} = 0$ -- correctly reproducing $K_{[1]}$.

\bigskip

On \textbf{layer 2} there are two polynomials  $P_{[2]}$ and $P_{[1, 1]}$, associated with partitions $[2]$ and $[1,1]$ respectfully, and since partitions appear explicitly in the sum in \eqref{eq:single-system-gauss}, we calculate them separately:
\[
P_{[2]} = D_{[2]} p_1^2 + C_{[2]} p_2 + B_{[2]} p_1 + A_{[2]}, \ \Rightarrow
    \left\{ \begin{aligned} 
  H_1 \cdot P_{[2]}= 2p_1 (D_{[2]} N + C_{[2]}) + N B_{[2]}  = (N+1) P_{[1]} \\
  H_2 \cdot P_{[2]}= 2N (D_{[2]} + C_{[2]} N) = N(N+1)P_{[0]}
\end{aligned} \right.
\]
Coefficients in front of each $p_k$-monomial must vanish separately, so we obtain the following system:
\[
\left\{ \begin{aligned}
    \{H_1 \ ; \ 1\}: \; \; N B_{[2]} = (N+1)A_{[1]} \\
    \{H_1 \ ; \ p_1\}: \; \; 2(D_{[2]} N + C_{[2]})  = N+1 \\
    \{H_2 \ ; \ 1\}: \; \; 2N(D_{[2]} + C_{[2]} N) = N(N+1)
\end{aligned} \right.
,\ \Rightarrow
    \left\{ \begin{aligned} 
  B_{[2]}  = \frac{N+1}{N}A_{[1]} \\
  D_{[2]}  = C_{[2]}  = \frac{1}{2} \\
\end{aligned} \right.
\]
Therefore we get $P_{[2]}=S_{[2]} + P'_{[2]}(A_{[1]}, \ p_1)+A_{[2]}$ where $P'_{[2]} = B_{[2]} p_1 = \frac{N+1}{N}A_{[1]} p_1$ meaning it inherited free parameters from the previous layer and did not add any new ones. The only new degree of freedom is free parameter $A_{[2]}$. Again, imposing \eqref{eq:gauss-dual-single-equation} for $\Delta = [2]$ and remembering that $A_{[1]} = 0$ we get $A_{[2]} = N(N+1) $ meaning $P_{[2]} = S_{[2]} +  N(N+1) =  K_{[2]}$
-- the correct distinguished solution is reproduced.
\newline
\[
P_{[1,1]} = D_{[1,1]} p_1^2 + C_{[1,1]} p_2 + B_{[1,1]} p_1 + A_{[1,1]}, \ \Rightarrow
    \left\{ \begin{aligned} 
  H_1 \cdot P_{[1,1]}= 2p_1 (N D_{[1,1]} + C_{[1,1]}) + N B_{[1,1]}  = (N-1) P_{[1]} \\
  H_2 \cdot P_{[1,1]}= 2N (D_{[1,1]} + N C_{[1,1]}) = -N(N-1)P_{[0]}
\end{aligned} \right.
,\ \Rightarrow
\]
\[
\Rightarrow
    \left\{ \begin{aligned}
    \{H_1 \ ; \ 1\}: \; \; N B_{[1,1]} = (N-1)A_{[1]} \\
    \{H_1 \ ; \ p_1\}: \; \; 2(N D_{[1,1]} + C_{[1,1]})  = N-1 \\
    \{H_2 \ ; \ 1\}: \; \; 2N(D_{[1,1]} + N C_{[1,1]}) = -N(N-1)
\end{aligned} \right.
,\ \Rightarrow
    \left\{ \begin{aligned} 
  B_{[1,1]}  = \frac{N-1}{N}A_{[1]} \\
  D_{[1,1]}  = -C_{[1,1]}  = \frac{1}{2} \\
\end{aligned} \right.
\]

Therefore we get $P_{[1, 1]}=S_{[1, 1]} + P'_{[1,1]}(A_{[1]}, \ p_1)+A_{[1, 1]}$. Similarly to previous case $P'_{[1, 1]} = B_{[1, 1]}p_1 = \frac{N-1}{N} A_{[1]} p_1$. Once again there are no new degrees of freedom except for a single parameter $A_{[1,1]}$. We then solve \eqref{eq:gauss-dual-single-equation} for $\Delta = [1, 1]$ and since $A_{[1]} = 0$ we get $A_{[1, 1]} = -N(N-1) $ meaning $P_{[1,1]} = S_{[1, 1]} -  N(N-1) = K_{[1,1]}$.

\bigskip

Prolonged calculations for \textbf{layer 3} are in a similar vein and can be seen in \textbf{Appendix A}. As a result, following patterns emerge:

\begin{enumerate}
\item For every partition $\Delta$: coefficients of $P_\Delta$ are defined by those in polynomials for smaller partitions except for exactly one, constant term $A_\Delta$ serving as an additional degree of freedom. Rigorous proof of this can be found in \textbf{Appendix B}.
\item All polynomials $P_\Delta$ have Schur polynomials $S_\Delta$ as their highest degree part.
\item If we additionally ask for \eqref{eq:gauss-dual-single-equation} as a constraint, all polynomials become uniquely defined, specifically as $P_\Delta = K_\Delta$. Proof for the general case can be found in \textbf{Appendix B}).
\end{enumerate}

We also note that generic solution $P_{\Delta}$ cannot be acquired (without additional constraints on constant terms $A_{\Delta}$) from Schur polynomials using linear transforms $p_k \rightarrow p_k' = p_k + x_k$. Even for $P_{[2]}$ and $P_{[1, 1]}$, if $p_1 \rightarrow p_1' = p_1 + A_{[1]}$ (so that $P_{[1]}(p) = S_{[1]}(p')$) and $p_2 \rightarrow p_2' = p_2 + X$: 
\[S_{[2]}(p') = \frac{1}{2}(p_2 + p_1^2) + \frac{X+A_{[1]}^2 + 2A_{[1]} p_1}{2} = P_{[2]}, \Rightarrow \frac{N+1}{N}A_{[1]} p_1 + A_{[2]} = A_{[1]} p_1 + \frac{X+A_{[1]}^2}{2}, \Rightarrow
\]
\[
\Rightarrow X = \frac{2A_{[1]} p_1}{N} + 2A_{[2]} - A_{[1]}^2
\]

\[S_{[1, 1]}(p') = \frac{1}{2}(p_1^2 - p_2) + \frac{A_{[1]}^2 + 2A_{[1]} p_1 - X}{2} = P_{[1,1]}, \Rightarrow \frac{N-1}{N}A_{[1]} p_1 + A_{[1,1]} = A_{[1]} p_1 + \frac{A_{[1]}^2 - X}{2}, \Rightarrow
\]
\[
\Rightarrow X = \frac{2A_{[1]} p_1}{N} - 2A_{[1,1]} + A_{[1]}^2
\]
Meaning $A_{[1]}^2 = A_{[1,1]} + A_{[2]}$ needs to be true which obviously is not true in general.

\bigskip

We now proceed with looking into analogs of augmented system \eqref{eq:gauss-dual-single-equation} + \eqref{eq:single-system-gauss} in other models, specifically WLZZ-models, in attempt to ascertain if existence of a unique solution continues to hold true.

\section{Generalization for WLZZ models} \label{sec:wlzz}
In this section we will be using previously described method to define polynomials $K_n^{(m)}(\Delta)$ for the positive branch of WLZZ matrix models constructed using Hamiltonians \footnote{For example, in $m=1$ those commutative sets give rise to the rational
Calogero Hamiltonians at the free fermion point \cite{paper:MM-int-syst-WLZZ}} $H_{\pm n}^{(m)}$ in place of $W_{\pm 2} = H_{\pm 2}^{(1)}$ \cite{paper:MM-int-syst-WLZZ}. Here we  are interested in, firstly, whether solution of augmented system remains being unique and, secondly, whether suitable analogs of orthogonality \eqref{eq:diagonality-gauss} and strict superintegrability \eqref{eq:strict-superint-gauss} are present. To both questions the answer is positive as we show below.

\bigskip

The superintegrability for WLZZ partition function on ray $(n, m)$ reads
\begin{align} \label{eq:partition-func-WLZZ}
Z_n^{(m)} = \underset{\Delta}{\sum} S_{\Delta}(p_k) \cdot S_{\Delta} (\delta_{k, n}) \cdot \Sbracket{\eta_\Delta(N)}^m
\end{align}
The corresponding Single equation is
\begin{align} \label{eq:gen-single-equation}
(D - H^{(m)}_{-n}) \cdot Z^{(m)}_n = 0,
\end{align}
therefore its corresponding dual equation reads
\begin{align} \label{eq:gen-dual-single-equation}
(D + H^{(m)}_n - |\Delta|) \cdot K_n^{(m)}(\Delta) = 0
\end{align}
Implementing our additional method, we suppose following system of "ladder" equations holds
\begin{align} \label{eq:single-system-gen}
\forall k: \ H^{(m)}_k \cdot K_n^{(m)}(\Delta)=\underset{\Delta' = \Delta - (k \ hook)}{\sum}  (-1)^{||k \ hook||} \cdot \Rbracket{\frac{\eta_\Delta (N)}{\eta_{\Delta'} (N)}}^m \cdot  K_n^{(m)}(\Delta')
\end{align}
One can observe with direct calculations that $K_n^{(m)}(\Delta)$ are indeed uniquely defined:
\begin{align} \label{eq:hermite-poly-gen}
 K^{(m)}_n(\Delta) = \underset{\Delta' \subseteq \Delta}{\sum} S_{\Delta'} (p_k) S_{\Delta / \Delta'}(\delta_{k,n}) \Sbracket{\frac{\eta_{\Delta}(N)}{\eta_{\Delta'}(N)}}^m
\end{align}
Furthermore, their double averages are still diagonal:
\begin{align}\label{eq:diagonality-gen}
<K_n^{(m)}(\Delta) \cdot K_n^{(m)}(\Delta')>_n^{(m)} \ = \ S_{\Delta}(N) \delta_{\Delta, \Delta'}
\end{align}
And they still fulfill strict superintegrability condition that can be straightforwardly generalized from Gaussian case:
\begin{align}\label{eq:strict-superint-gen}
\boxed{
<S_{\Delta} \cdot K^{(m)}_n({\Delta')}>_n^{(m)} \ = \ S_{\Delta / \Delta'}(\delta_{k, n}) [\eta_{\Delta}(N)]^m
}
\end{align}
Which is an expected consequence of (ordinary) superintegrability formula in WLZZ-models:
\begin{align}
<S_{\Delta}>_n^{(m)} \ = \ S_{\Delta} (\delta_{k, n}) \cdot \Sbracket{\eta_{\Delta}(N)}^m
\end{align}

We note that the formula \eqref{eq:strict-superint-gen} is direct and straightforward generalization from the Gaussian case -- in contrast to cases of monomial matrix models
\cite{paper:CMPT-monomial-strict-superint} where strict superintegrability was so far observed only for certain subclasses of partitions. Moreover, it featured peculiar permutation operator
$\pi$ on the space of partition $r$-quotients. Here no such subtleties are necessary and everything simply "works".
This is probably due to the fact that WLZZ-models are related to HGMM within the framework of ordinary $W_{1+\infty}$ (Yangian) algebra, while monomial matrix models are naturally embedded into Yangian/$W_{1+\infty}$ algebra for $gl_r$. This direction, however, will be studied elsewhere.

\subsection{WLZZ-model Miwa deformation}
Now that we have the strict superintegrability formula \eqref{eq:strict-superint-gen} we can repeat the process from \cite{paper:MMPSh-miwa} to reach similar results for Miwa-deformed WLZZ-models.

\bigskip

The integral representation in the general $(n,m)-$WLZZ case is more involved -- it is
$m+1-$matrix model. However, our algebraic approach allows us to address this case with no additional complications
\begin{align} \label{eq:partition-func-in-Miwa-deformed-gen}
Z_{n, \pi_k}^{(m)} = & \ \underset{\Delta}{\sum} S_{\Delta}(p_k + \pi_k) \cdot S_{\Delta} (\delta_{k, n}) \cdot \Sbracket{\eta_\Delta(N)}^m = \underset{\Delta, \Delta'}{\sum} S_{\Delta'}(p_k) \cdot S_{\Delta / \Delta'} (\pi_k) \cdot S_{\Delta} (\delta_{k, n}) \cdot \Sbracket{\eta_\Delta(N)}^m
\\ \notag
= & \ \Bigg{|}_{\pi_k=\sum_{i=1}^M z_i^k} \ \int DX_0\dots DX_m
\cdot \prod_{i=0}^{m-1}\exp(\Tr X_i X_{i+1})
\cdot \exp(-\frac{1}{n} \Tr X_m^n)
\cdot \frac{\exp{\Rbracket{\overset{\infty}{\underset{k=1}{\sum}} \frac{p_k}{k} \Tr X_0^k}}}{\prod_{i=1}^M \text{det}\left(1 - z_i X_0\right)}
\end{align}
Once again if we can conveniently calculate average of any $K_n^{(m)}(\Delta)$ polynomial
as Borel transform in the Miwa variables
\begin{empheq}[box=\fbox]{align} \label{eq:averages-in-Miwa-deformed-WLZZ}
<K_n^{(m)}(\Delta')>_{n, <s>}^{(m)} = & \ \underset{\Delta}{\sum} S_{\Delta}(z_a) <S_{\Delta} \cdot K_n^{(m)}(\Delta')>_{n}^{(m)} = 
\underset{\Delta}{\sum} S_{\Delta}(z_a) \cdot S_{\Delta / \Delta'} (\delta_{k,n}) \cdot \Sbracket{\eta_\Delta(N)}^m
\\ \notag
= & \  \mathcal{B}_N^m \Sbracket{\underset{\Delta}{\sum} S_{\Delta}(z_a) \cdot S_{\Delta / \Delta'}(\delta_{k,n})} = \mathcal{B}_N^m \Sbracket{S_{\Delta'}\{\underline{z}\}
e^{\frac{1}{2}\sum_a z_a^n}}
\end{empheq}

This along with specific construction formulas \eqref{eq:hermite-poly-gen} of polynomials $K_n^{(m)}(\Delta)$ is the main result of this paper.

\bigskip

Note that as an easy consequence of \eqref{eq:single-system-gen} one may write $K_n^{(m)}(\Delta)$ polynomials in exponential form as
\begin{align} \label{eq:hermite-poly-w-gen}
  K_n^{(m)}(\Delta) = \Rbracket{\exp{\frac{H^{(m)}_n}{n}}} \cdot S_\Delta
\end{align}

We also note that generalization from the case of just one point on a given ray
(potential $-\frac{1}{n}\Tr X_m^n$) to the case of sum of Hamiltonians on \textit{the same ray}
(potential $-\sum_n \frac{g_n}{n}\Tr X_m^n$) is straightforward and just involves changing
$\delta_{k,n} \rightarrow g_n$ in our formula \eqref{eq:averages-in-Miwa-deformed-WLZZ}.
The case of cones of Hamiltonians seems to be more involved and will be treated elsewhere.

\section{Conclusion}\label{sec:conclusion}
In this paper we considered possibility of Miwa-deformation for WLZZ matrix models. We argued that once suitable generalization of multivariate orthogonal polynomials $K_{\Delta}$ is found,
the derivation essentially parallels that for Gaussian case and results in Miwa-deformed superintegrable formula \eqref{eq:partition-func-in-Miwa-deformed-gen}. The two main ingredients used to construct this suitable $K_{\Delta}$ generalization are the dual Single equation and 
the "ladder" equations obtained from action of higher Hamiltonians.
Together they uniquely define polynomials \eqref{eq:hermite-poly-gen}.

\bigskip

There are several directions in which research can be continued:
\begin{itemize}
\item How to generalize this approach to monomial matrix models \cite{paper:P-towards-mixed-phase, paper:CMPT-monomial-strict-superint}? In particular, this generalization should provide natural explanation
to so far mysterious shuffle operation $\pi$ (see (10) in
\cite{paper:CMPT-monomial-strict-superint}).
\item Does our construction admit generalization to $\beta$ (Yangian) and $(q, t)$ (DIM) deformations? In particular, what is the interplay with existing superintegrable formulas in terms of Uglov polynomials \cite{paper:MM-superint-uglov}?
\item In this paper only WLZZ-models constructed from the positive branch of $W_{1+\infty}$ algebra were considered in detail. While negative branch case seems analogous, still, it should in principle be thouroughly worked out.
\item It may be interesting to consider cones in $W_{1+\infty}$ algebra (linear combinations of $H^{(m)}_n$ with different parameters $m$ and constant $n$) as replacements to regular Hamiltonian rays.
\item Superintegrable basis for WLZZ-models is known and provided by Schur polynomials $S_{\Delta}$. Our results for $K_{\Delta}$ correctly agree with this. Now the next step is to apply our recipe to a model with unknown superintegrable basis and to see if any interesting results come out of it -- first natural candidate for this approach being the celebrated Kontsevich model.

\end{itemize}

The work is continuing in these directions.

\section*{Acknowledgements}
We would like to thank A.Mironov, A.Morozov, L.Bishler, A.Oreshina and N.Tselousov for stimulating discussions.

\bigskip

A.P. gratefully acknowledges support from the Ministry of Science and Higher Education of the Russian Federation (agreement no.075-03-2025-662).

\section*{Appendix A}
\[
P_{[3]} = J_{[3]} p_3 + G_{[3]} p_2 p_1 + F_{[3]} p_1^3 +   D_{[3]} p_1^2 + C_{[3]} p_2 + B_{[3]} p_1 + A_{[3]}, \ \Rightarrow \]
\[
\Rightarrow
    \left\{ \begin{aligned} 
  H_1 \cdot P_{[3]}= 3p_2 J_{[3]} + (Np_2 + 2p_1^2)G_{[3]} + 3N p_1^2 F_{[3]} + 2p_1 (ND_{[3]} + C_{[3]}) + N B_{[3]}  = (N+2) P_{[2]} \\
  H_2 \cdot P_{[3]}= 6N p_1 J_{[3]} + 2 (2+N^2) p_1 G_{[3]} + 6Np_1 F_{[3]} + 2N (D_{[3]} + NC_{[3]}) = (N+2)(N+1) P_{[1]} \\
  H_3 \cdot P_{[3]}= 3(N^3+N) J_{[3]} + 6N^2 G_{[3]} + 6N F_{[3]} = (N+2)(N+1)N P_{[0]}
\end{aligned} \right.
,\ \Rightarrow
\]
\[
\Rightarrow
    \left\{ \begin{aligned}
    \{H_1 \ ; \ 1\}: \; \; N B_{[3]} = (N+2)A_{[2]} \\
    \{H_1 \ ; \ p_1\}: \; \; 2(N D_{[3]} + C_{[3]}) = (N+2)B_{[2]} \\
    \{H_1 \ ; \ p_2\}: \; \; 3 J_{[3]} + N G_{[3]} = (N+2)C_{[2]} = \frac{N+2}{2} \\
    \{H_1 \ ; \ p_1^2\}: \; \; 2 G_{[3]} + 3N F_{[3]}  = (N+2) D_{[2]} = \frac{N+2}{2} \\
    \{H_2 \ ; \ 1\}: \; \; 2N(D_{[3]} + NC_{[3]}) = (N+2)(N+1)A_{[1]} \\
    \{H_2 \ ; \ p_1\}: \; \; 6N(J_{[3]} + F_{[3]}) + 2(2+N^2)  G_{[3]} = (N+2)(N+1)B_{[1]} \\
    \{H_3 \ ; \ 1\}: \; \; 3(N^3+N) J_{[3]} + 6N^2 G_{[3]} + 6N F_{[3]} = (N+2)(N+1)N
\end{aligned} \right.
\ \Rightarrow
    \left\{ \begin{aligned} 
  B_{[3]}  = \frac{N+2}{N}A_{[2]} \\
  D_{[3]}  = NC_{[3]}  = \frac{N+2}{2N}A_{[1]} \\ J_{[3]} = \frac{1}{3} \\ G_{[3]} = \frac{1}{2} \\ F_{[3]} = \frac{1}{6}
\end{aligned} \right.
\]
Therefore we get $P_{[3]}=S_{[3]} + P'_{[3]}(A_{[2]}, \ A_{[1,1]}, \ A_{[1]}, \  p_2, \ p_1)+A_{[3]}$. The only additional free parameter is $A_{[3]}$. We then solve \eqref{eq:gauss-dual-single-equation} for $\Delta = [3]$ and get $A_{[3]} = 0$ meaning $P_{[3]} = S_{[3]} + (N+1)(N+2) p_1=  K_{[3]}$.

\bigskip

\[
P_{[2,1]} = J_{[2,1]} p_3 + G_{[2,1]} p_2 p_1 + F_{[2,1]} p_1^3 + D_{[2,1]} p_1^2 + C_{[2,1]} p_2 + B_{[2,1]} p_1 + A_{[2,1]}, \ \Rightarrow \]
\[
\Rightarrow
    \left\{ \begin{aligned} 
  H_1 \cdot P_{[2,1]}= 3p_2 J_{[2,1]} + (Np_2 + 2p_1^2)G_{[2,1]} + 3N p_1^2 F_{[2, 1]} + 2p_1 (ND_{[2,1]} + C_{[2,1]}) + N B_{[2,1]}  = (N-1) P_{[2]} + (N+1)P_{[1,1]} \\
  H_2 \cdot P_{[2,1]}= 6N p_1 J_{[2,1]} + 2 (2+N^2) p_1 G_{[2,1]} + 6Np_1 F_{[2,1]} + 2N (D_{[2,1]} + NC_{[2,1]}) = 0 \\
  H_3 \cdot P_{[2,1]}= 3(N^3+N) J_{[2,1]} + 6N^2 G_{[2,1]} + 6N F_{[2,1]} = -(N-1)N(N+1)P_{[0]}
\end{aligned} \right.
,\ \Rightarrow
\]
\[
\Rightarrow
    \left\{ \begin{aligned}
    \{H_1 \ ; \ 1\}: \; \; N B_{[2,1]} = (N-1)A_{[2]} + (N+1)A_{[1,1]} \\
    \{H_1 \ ; \ p_1\}: \; \; 2(N D_{[2,1]} + C_{[2,1]}) = (N-1)B_{[2]} + (N+1)B_{[1,1]} \\
    \{H_1 \ ; \ p_2\}: \; \; 3 J_{[2,1]} + N G_{[2,1]} = (N-1)C_{[2]} + (N+1)C_{[1,1]} = -1 \\
    \{H_1 \ ; \ p_1^2\}: \; \; 2 G_{[2,1]} + 3N F_{[2,1]}  = (N-1) D_{[2]} + (N+1)D_{[1,1]} = N \\
    \{H_2 \ ; \ 1\}: \; \; 2N(D_{[2,1]} + NC_{[2,1]}) = 0 \\
    \{H_2 \ ; \ p_1\}: \; \; 6N(J_{[2,1]} + F_{[2,1]}) + 2(2+N^2)  G_{[2,1]} = 0 \\
    \{H_3 \ ; \ 1\}: \; \; 3(N^3+N) J_{[2,1]} + 6N^2 G_{[2,1]} + 6N F_{[2,1]} = -(N-1)N(N+1)
\end{aligned} \right.
\ \Rightarrow
    \left\{ \begin{aligned} 
  B_{[2,1]}  = \frac{N-1}{N}A_{[2]} + \frac{N+1}{N}A_{[1,1]} \\
  D_{[2,1]}  = -NC_{[2,1]}  = 2 A_{[1]} \\ J_{[2,1]} = -\frac{1}{3} \\ G_{[2,1]} = 0 \\ F_{[2,1]} = \frac{1}{3}
\end{aligned} \right.
\]
Therefore we get $P_{[2, 1]}=S_{[2, 1]} + P'_{[2,1]}(A_{[2]}, \ A_{[1,1]}, \ A_{[1]}, \  p_2, \ p_1)+A_{[2, 1]}$. The only additional free parameter is $A_{[2, 1]}$. We then solve \eqref{eq:gauss-dual-single-equation} for $\Delta = [2, 1]$ and get $A_{[2, 1]} = 0$ meaning $P_{[2, 1]} = S_{[2, 1]} =  K_{[2, 1]}$.

\bigskip

\[
P_{[1, 1,1]} = J_{[1,1,1]} p_3 + G_{[1,1,1]} p_2 p_1 + F_{[1,1,1]} p_1^3 +   D_{[1,1,1]} p_1^2 + C_{[1,1,1]} p_2 + B_{[1,1,1]} p_1 + A_{[1,1,1]}, \ \Rightarrow \]
\[
\Rightarrow
    \left\{ \begin{aligned} 
  H_1 \cdot P_{[1,1,1]}= 3p_2 J_{[1,1,1]} + (Np_2 + 2p_1^2)G_{[1,1,1]} + 3N p_1^2 F_{[1,1,1]} + 2p_1 (N D_{[1,1,1]} + C_{[1,1,1]}) + N B_{[1,1,1]}  = (N-2)P_{[1,1]} \\
  H_2 \cdot P_{[1,1,1]}= 6N p_1 J_{[1,1,1]} + 2 (2+N^2) p_1 G_{[1,1,1]} + 6Np_1 F_{[1,1,1]} + 2N (D_{[1,1,1]} + N C_{[1,1,1]}) = -(N-2)(N-1) P_{[1]} \\
  H_3 \cdot P_{[1,1,1]}= 3(N^3+N) J_{[1,1,1]} + 6N^2 G_{[1,1,1]} + 6N F_{[1,1,1]} = (N-2)(N-1)N P_{[0]}
\end{aligned} \right.
,\ \Rightarrow
\]
\[
\Rightarrow
    \left\{ \begin{aligned}
    \{H_1 \ ; \ 1\}: \; \; N B_{[1,1,1]} = (N-2)A_{[1,1]} \\
    \{H_1 \ ; \ p_1\}: \; \; 2(N D_{[1,1,1]} + C_{[1,1,1]}) = (N-2)B_{[1,1]} \\
    \{H_1 \ ; \ p_2\}: \; \; 3 J_{[1,1,1]} + N G_{[1,1,1]} = (N-2)C_{[1,1]} = -\frac{N-2}{2} \\
    \{H_1 \ ; \ p_1^2\}: \; \; 2 G_{[1,1,1]} + 3N F_{[1,1,1]}  = (N-2)D_{[1,1]} = \frac{N-2}{2} \\
    \{H_2 \ ; \ 1\}: \; \; 2N(D_{[1,1,1]} + C_{[1,1,1]} N) = -(N-2)(N-1)A_{[1]} \\
    \{H_2 \ ; \ p_1\}: \; \; 6N(J_{[1,1,1]} + F_{[1,1,1]}) + 2(2+N^2)  G_{[1,1,1]} = -(N-2)(N-1)B_{[1]} \\
    \{H_3 \ ; \ 1\}: \; \; 3(N^3+N) J_{[1,1,1]} + 6N^2 G_{[1,1,1]} + 6N F_{[1,1,1]} = (N-2)(N-1)N
\end{aligned} \right.
\ \Rightarrow
    \left\{ \begin{aligned} 
  B_{[1,1,1]}  = \frac{N-2}{N}A_{[1,1]} \\
  D_{[1,1,1]}  = -C_{[1,1,1]}  =\frac{N-2}{2N} A_{[1]} \\ J_{[2,1]} = \frac{1}{3} \\ G_{[2,1]} = -\frac{1}{2} \\ F_{[2,1]} = \frac{1}{6}
\end{aligned} \right.
\]
Therefore we get $P_{[1,1,1]}=S_{[1,1,1]} + P'_{[1,1,1]}(A_{[2]}, \ A_{[1,1]}, \ A_{[1]}, \  p_2, \ p_1)+A_{[1,1,1]}$. The only additional free parameter is $A_{[1, 1, 1]}$. We then solve \eqref{eq:gauss-dual-single-equation} for $\Delta = [1, 1, 1]$ and get $A_{[1, 1, 1]} = 0$ meaning $P_{[1, 1, 1]} = S_{[1, 1, 1]} - (N-1)(N-2)p_1 =  K_{[1, 1, 1]}$.

\section*{Appendix B}
In this appendix we provide proof of uniqueness for $K_\Delta$ for Gaussian Hermitian Matrix Model. Let us first formally separate arbitrary polynomial into "grading layers" $\mathcal{K}_l(\Delta)$ and index them with whole numbers $l$ indicating difference between gradings of layer and partition $\Delta$:
\begin{align}
K_\Delta = \underset{l}{\sum}\mathcal{K}_l(\Delta)
\end{align}
\begin{align}
\mathcal{K}_l(\Delta) = \underset{\Delta' : |\Delta'| = |\Delta| + l}{\sum} C_{\Delta'} \cdot \underset{ k \in \Delta'} {\prod} p_{k}
\end{align}

It is easy to notice that equations \eqref{eq:single-system-gauss} are equivalent to system of same equations written for all $\mathcal{K}_l(\Delta)$ instead of $K_\Delta$.

\begin{enumerate}
\item We start by checking $l=0$ and proving $\mathcal{K}_0(\Delta) = S_\Delta$ to be a unique solution of \eqref{eq:single-system-gauss} in grading layer zero:
If $\mathcal{K}_0([0]) = \alpha$ and $\mathcal{K}_0([1]) = A p_1$, then according to \eqref{eq:single-system-gauss} $NA=N\alpha$, so $\mathcal{K}_0([1]) = \alpha p_1 = \alpha S_{[1]}$ is defined uniquely. For further proof we assume $\alpha=1$.
Next we use induction and prove that if $\forall \Delta:|\Delta|=m-1$ solution $\mathcal{K}_0(\Delta) = S_\Delta$ is unique, then the same would be true for any $\Delta$ with $|\Delta|=m$:

First, let us denote all partitions $\Delta'$ with $|\Delta'| = n$ as $I_n$ and also index partitions in each of them: $1 \leq j \leq |I_n|: \ \Delta_{j} \in I_n$. Next, we construct a system of linear equations that will be equivalent to \eqref{eq:single-system-gauss}: we write $K_\Delta$ as $\underset{\Delta' \in I_{|\Delta|}}{\sum} x_{\Delta'}(\Delta) \cdot p_{\Delta'}$ \footnote{Here $p_{\Delta'}$ means $\underset{i \in \Delta'}{\prod} p_i$} and get equations on variables $x_{\Delta'}$:
\begin{align}
M X = R, \ \ X = \begin{pmatrix} x_{[\underset{n}{\underbrace{1, 1, \ldots, 1}}}] \\ \vdots \\ x_{[n-1, 1]} \\ x_{[n]}\end{pmatrix} 
\end{align}
Here $M$ is a $|I_{|\Delta|}| \times |\Delta|$ matrix. Its element $(i, j)$ represents the way $H_i$ acts on $p_{\Delta_j}$.
In case $n=2$, for example, $M= \begin{pmatrix} 2N(p_1) & 2(p_1) \\ 2N & 2N^2 \end{pmatrix}$. Right-hand side $R$ for any specific $\Delta$ can also be gleamed from system \eqref{eq:single-system-gauss}: element $i$ should consist of all the terms of grading layer $|\Delta| - i$ in original right-hand side. For example, $R_{[1,1]} = \begin{pmatrix} (N-1)(p_1) \\ -N(N-1)\end{pmatrix}$ and  $R_{[2]} = \begin{pmatrix} (N+1)(p_1) \\ N(N+1)\end{pmatrix}$. Any resulting lynear system's solution can be maped equivalently to layer zero of a system \eqref{eq:single-system-gauss} solution followng $\mathcal{K}_0(\Delta_j)= \underset{\Delta' \in I_{n}}{\sum} x_{\Delta'}(\Delta_j) \cdot p_{\Delta'}$. This means all that's left is to prove the system has a unique solution for $\mathcal{K}_0(\Delta_j)$ to be a unique layer zero polynomial solution of \eqref{eq:single-system-gauss} with a given $\Delta_j$. An equivalent to this statement would be "There cannot be a non-trivial solution $Y$ resulting in MY=0" meaning there are no non-trivial polynomials $P_Y(\Delta) =\underset{\Delta' \in I_{|\Delta|}}{\sum} y_{\Delta'} \cdot p_{\Delta'} \ \hookrightarrow \ \forall k \in \mathbb{N}: H_k(P_Y)=0$. If we consider it wrong and assume at least one such polynomial exists, it must also fulfill property $\forall \Delta_{j} \in J_{|\Delta|}: H_{\Delta_j}(P_Y) = 0$ where $H_{\Delta} = \underset{i \in \Delta}{\prod} H_i$ and therefore we can write a new matrix $M'$ by acting with combination of Hamiltonians $H_{\Delta_i}$ on $p_{\Delta_j}$ to get element $(i, j)$. That way $M'$ is a $|I_{|\Delta|}| \times |I_{|\Delta|}|$ matrix with no element depending on any $p_k$, only on $N$. Meaning our $Y$ has to remain a non-trivial solution to
\begin{align}\label{eq:matrix-mult-2}
M' Y = 0
\end{align}
Where we notice that $M'$ has following structure: for every row there exists exactly one element that has the highest power of $N$ which is $|\Delta|$ (it is in the column being multiplied by $y_{\Delta'}$ if the row itself correlates to $H_{\Delta'}$ being previously used) while all other elements of the same row consist only of sums of strictly lower powers of $N$. That means that $\forall \Delta_j \in J_{|\Delta|}$ there exists a row where in the element of $M'Y$ in front of $N^{|\Delta|}$ would be just $y_{\Delta_j}$, meaning for \eqref{eq:matrix-mult-2} to be true it we need $\forall \Delta_j \in I_{|\Delta|} : \ y_{\Delta_j} = 0$. Meaning $Y$ is trivial. Contradiction.

\item For $l>0$: if $\exists \ \Delta: \mathcal{K}_l(\Delta) \neq 0$, then we can find the smallest $i$ so that $\exists \Delta' : |\Delta'|=i  \ (>0)$ and $\mathcal{K}_l(\Delta') \neq 0$, but $\forall j<i:|\Delta|=j \ \Rightarrow \ \mathcal{K}_l(\Delta)=0$. Then for aforementioned $\Delta'$ from \eqref{eq:single-system-gauss} we have $1 \leq k \leq   |\Delta'| : H_k (K_{\Delta'}) =0$. We also have general notion that $\forall k > |\Delta'| : H_k (K_{\Delta'}) =0$ meaning $\forall k \geq 1:H_k(\mathcal{K}_l(\Delta'))=0$, meaning (as was found while considering an $i=0$ case) $\mathcal{K}_l(\Delta')=0$. Contradiction.

\item For $l<0$: what will happen if we add additional constraints \eqref{eq:gauss-dual-single-equation}? We can notice that for this equation to be true in grading layer terms following has to be true $\forall l$:
\begin{equation}
    \begin{split}
|\Delta_k|\mathcal{K}_l(\Delta_k)  - W_2 \cdot \mathcal{K}_{l+2}(\Delta_k) = D \cdot \mathcal{K}_{l}(\Delta_k) = \\
= (|\Delta_k|+l) \mathcal{K}_{l}(\Delta_k) \Rightarrow \mathcal{K}_{l}(\Delta_k) = \frac{W_2 \cdot \mathcal{K}_{l+2}(\Delta_k)}{-l}
    \end{split}
\end{equation}
And since we already know $\forall \Delta: \mathcal{K}_0(\Delta)=S_\Delta$ and $\mathcal{K}_{\geq 1}(\Delta)=0$, all of $\mathcal{K}_l(\Delta)$ for every $l<0$ and every $\Delta$ can be uniquely defined. Specifically that way we get precisely the W-generating formula \eqref{eq:hermite-poly-w-gauss} from Schur polynomials.
\end{enumerate}

Now since all grading layers of all polynomials are uniquely defined, we can say that polynomials $K_{\Delta}$ themselves are uniquely defined solutions of equations \eqref{eq:single-system-gauss} with additional constraint \eqref{eq:gauss-dual-single-equation}.

\end{document}